\documentstyle[prc,aps,epsf]{revtex}
\begin{document}

\draft


\title{An Update on NA50 and LUCIFER}  

\author{D.~E.~Kahana$^{2}$ and S.~H.~Kahana$^{1}$}
 
\address{$^{1}$Physics Department, Brookhaven National Laboratory\\
   Upton, NY 11973, USA\\
   $^{2}$31 Pembrook Dr., Stony Brook, NY 11790, USA}
\date{\today}  
  
\maketitle  
  
\begin{abstract}
Recent analysis of both new (1998) and old (1995,1996) data obtained by the
NA50 Collaboration on J/$\psi$ suppression in Pb+Pb are examined in light of
our already existing calculations with the relativistic heavy ion simulation
LUCIFER. In particular we comment on the unexplained change of transverse
energy scale by NA50 and the apparent disappearance from the data of
discontinuities in the suppression with $E_t$.
\end{abstract}


\section{J/$\psi$ Suppression in `99} 
This brief note is presented to clear up any misinformation that may have
been created by the NA50 contribution to QM'99 \cite{Kluberg99} or to the
Proceedings of the International school of Nuclear Physics Erice, 17-25
September, 1998 \cite {Kluberg99a}.  The QM'99 document contains new data
from the NA50 collaboration, taken in late 1998 with a thinner target than
used previously. The total data sample is reanalysed and compared to a
variety of theoretical simulations. In at least one case this comparison was
not straightforwardly made \cite{Kahana98,Kahana99}. There are two
noteworthy features of the combined new data set. One is the striking absence
of the discontinuity in the $E_t$ spectrum which first appeared in the 1996
measurements \cite{NA50`97} for J/$\psi$ from Pb+Pb.  The second is the
very evident change in the $E_t$ scale between the present and several earlier
submissions. Not much is said in the QM'99 NA50 presentation by way of
explaining either of these changes, although presumably the `minimum-bias'
smoothing procedure, introduced in QM'99, accomplishes the removal of the
discontinuity.

The lack of such a singularity was of course anticipated in any theoretical 
calculation based on cascade-like simulations. Only theories proposing
some or other phase change have introduced such  singular behaviour, and it 
must be said, generally have done so in a rather {\it ad hoc} fashion. It is
in fact not clear to what extent discontinuities can persist in finite
systems, even when a `change of phase' is present in an infinite medium.

The change of scale for transverse energy was also anticipated in at least
one theoretical work \cite{Kahana98,Kahana99}. In the present authors'
discussion of J/$\psi$ suppression a scale factor was clearly referred to in
these two publications \cite{Kahana98,Kahana99}.  In Reference
\cite{Kahana99} this was done specifically in the caption to the figure
describing the comparison with NA50 for Pb+Pb at $158$ GeV (Figure 15 in
\cite{Kahana99}, Figure 1 below). At that point it was noted that
an $E_t$ scale factor had been introduced to reconcile the experimental and
theoretical spectra. Again in the earlier Reference \cite{Kahana98}, in the
caption to Figure 11, the reader is referred to the text for a discussion of
the $E_t$ scale, wherein it is stated that a scaling factor of $\sim 1.25$
was employed. Very clearly then, the present authors gave warning that the
Pb+Pb transverse energy scale achieved with LUCIFER was not in accord with 
that presented by the NA50 collaboration.

Reference was also made in both publications to private communications with
the NA50 collaboration. These communications led to the necessity of a scale
change and revealed that the `Collaboration' did not at that time actually
have good knowledge of the absolute $E_t$ scale. Indeed, a figure of $125$ GeV
for the end point was cited as a reasonable alternative to the heretofore
published value of 150-160 GeV \cite{NA50`97}. We are of course pleased if
the absolute $E_t$ is now better understood by NA50 and for the moment, at
least, is closer to our estimate.

In fact, the scale factor we used to compare the LUCIFER calculation with
experiment also took account of the seeming $5-10\%$ discrepancy in cutoff
between the full $E_t$ spectra of NA49 \cite{NA49} and LUCIFER (see Figure 13
in Reference \cite{Kahana99} which is Figure 2 here). The agreements between
simulations and both the NA49 inclusive meson and baryon spectra
\cite{Kahana98a} and this NA49 $E_t$ spectrum suggest an inconsistency with
the cutoff earlier quoted for NA50 data. The theoretical calculation acts as
an interpolation between experiments and predicted a cutoff transverse energy
nearer $120$ GeV than $150$. Thus the overall factor was close to $1.30$,
{\it i.\hskip -2pt e.} the factor between LUCIFER and NA50 '96 $E_t$ scales.

To further clear up any possible misapprehensions, we present here the
earlier Figures 13 and 15 from Reference \cite{Kahana99} (Figures 1 and 2
here) and a new Figure 3 comparing our $E_t$-unscaled spectrum for Pb+Pb with
one of the recent NA50 figures (Figure 57 in Kluberg QM'99). This NA50 plot
is apparently obtained by rebinning from more complete spectra, but the
overall effect is the same as graphing the totality of $E_t$ measurements. It
is clear that some of our last J/$\psi$ to Drell-Yan values, at the end of
the previous $E_t$ scaled calculation, appear slightly more suppressed. No
significance can be attached to this since any cacscade is necessarily an
inexact theory, to say the least, and our normalisatons are subject to some
error from taking a ratio to Drell-Yan, perhaps $7\%$ or less. It would have
been better to compare the theoretical survival probability directly to some
experimental estimate of this quantity. The calculated survival rates are
unchanged from our previous calculation. Of course none of these
normalisation problems attach to the comparison with minimum bias J/$\psi$
production, which in both simulation and experiment in principle use absolute
cross-sections. Our explanation of the anomalous suppression there
\cite{Kahana98,Kahana99} remains in place.

If anything, in this presentation of `unscaled $E_t$,' we have exaggerated a
small discrepancy at peripheral $E_t$, where unfortunately one does not
expect any unusual or `plasma-like' behaviour, and one must keep in mind the
NA50 caveat concerning their absolute $E_t$ scale. We must still conclude
that present deviations with NA50 do not justify any claims for startling
medium-based effects. One might well argue that the breakup of a small object
like the J/$\psi$ could never be ascribed to screening by a plasma. 
Ultimately, dissolution of the J/$\psi$ must result from gluon exchange
interactions between quarks initially in hadrons and in the $c \bar c$
preresonant pair. It is probably hard for the charmonium state to distinguish
between three quarks in a nucleon say, and the same three quarks somewhat
spread out as in a plasma. There is no true continuous medium which can
permeate the bound or preresonant charmonium state.

The cascade theory can hardly be called {\it ad hoc,} as it is described in
Reference \cite{Kluberg99a}. An attempt is simply made to incorporate as much
information as is known from the elementary hadronic data in a 
comprehensive multi-scattering formalism. In our comparison with inclusive
NA49 Pb+Pb spectra only a single intrinsic parameter of the model was
determined from ion-ion data, {\it i.\hskip -2pt e.} the formation time for
secondary mesons \cite{Kahana98a}, and that was obtained from the light
system S+S. The resulting good description of the NA49 Pb+Pb spectra surely
removes the theory from any {\it ad hoc} category. The same cannot be said
for the `Glauber' calculations which yield neither inclusive meson spectra
nor direct $E_t$ distributions, and which nevertheless were used to justify
the inability of standard theory to explain the suppression in Pb+Pb. In
particular, close to the correct number of produced mesons is achieved in the
LUCIFER simulation and thus the breakup of charmonium states by these
comovers is appropriately estimated, a  feature intimately tied to the
theory's correct evaluation of the total transverse energy.

For the purposes of calculating J/$\psi$ suppression, one requires other
cross-sections. Breakup of J/$\psi$ from its collisions with baryons is
determined from the nucleon-nucleus production data; breakup cross-sections
on mesons are essentially taken as $2/3$ of that on nucleons. We indicated
\cite{Kahana99} that breakup in meson-charmonium collisions mostly takes
place well above threshold, so the latter estimate is likely good. 

It is incumbent on those proposing the production of `plasma' in their
measurements to demonstrate a clear deviation with the normal `background' a
cascade provides. This will prove as necessary at RHIC as it was at the SPS.

\section{Acknowledgments}

The authors are grateful to Boris Kopeliovitch for several illuminating
discussions.  The present manuscript has been authored under US DOE grant
No. DE-AC02-98CH10866. One of us  (SHK) is pleased to acknowledge continuing
support from the Alexander Von Humbodt Foundation, Bonn Germany.

\begin{figure}
\vbox{\hbox to\hsize{\hfil
\epsfxsize=6.4truein\epsffile[24 85 577 736]{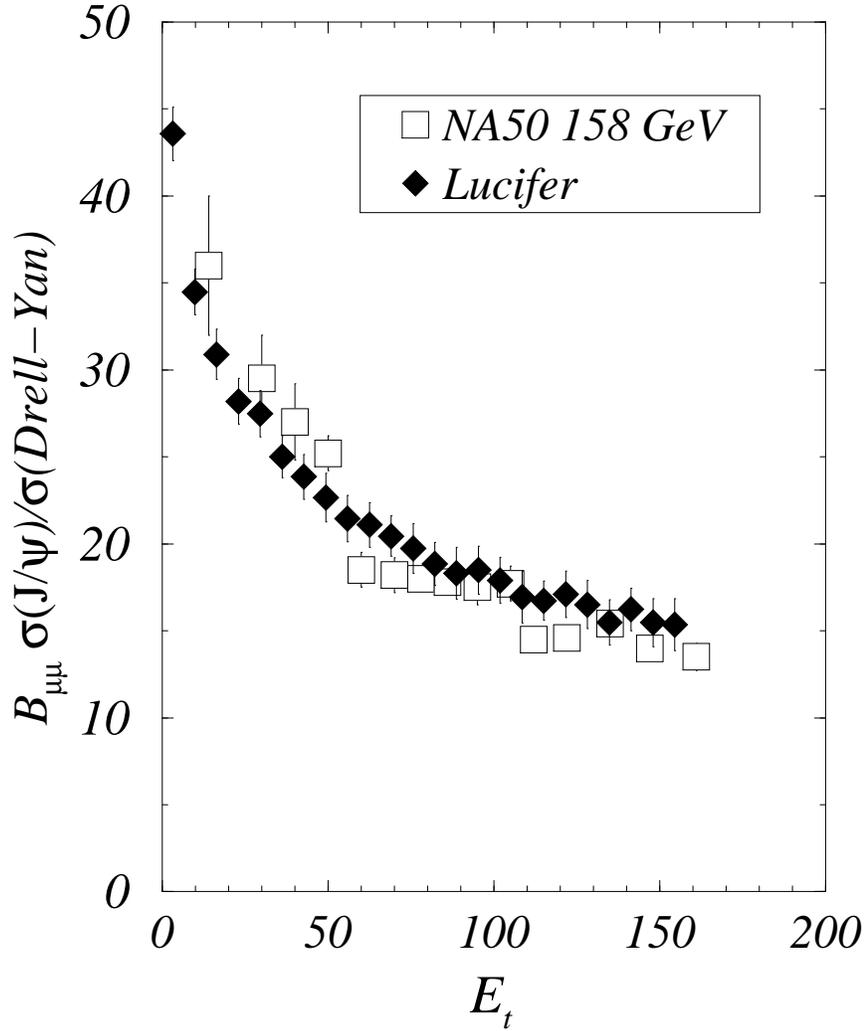}
\hfil}}
\caption[]{Ratio of J/$\psi$ to Drell-Yan in Pb+Pb. Comparison between the
cascade and NA50 (1996) neutral transverse energy dependence for
J/$\psi$. There are no discontinuities, of course, in the LUCIFER yields, but
the general shape is reproduced. The pseudorapidity range is here 1.1-2.3 and
a factor 1.2 used to normalize the theoretical energy scale[35]. The
experimental data was rescaled to 200 GeV/c by the experimentalists. The
theoretical normalisation (of J/$\psi$ to Drell-Yan) in both this comparison
and that for $S+U$ are subject to choices made for the elementary $pp$
values.}

\vspace*{10mm}
\label{fig:one}
\end{figure}
\clearpage

\begin{figure}
\vbox{\hbox to\hsize{\hfil
\epsfxsize=6.1truein\epsffile[24 85 577 736]{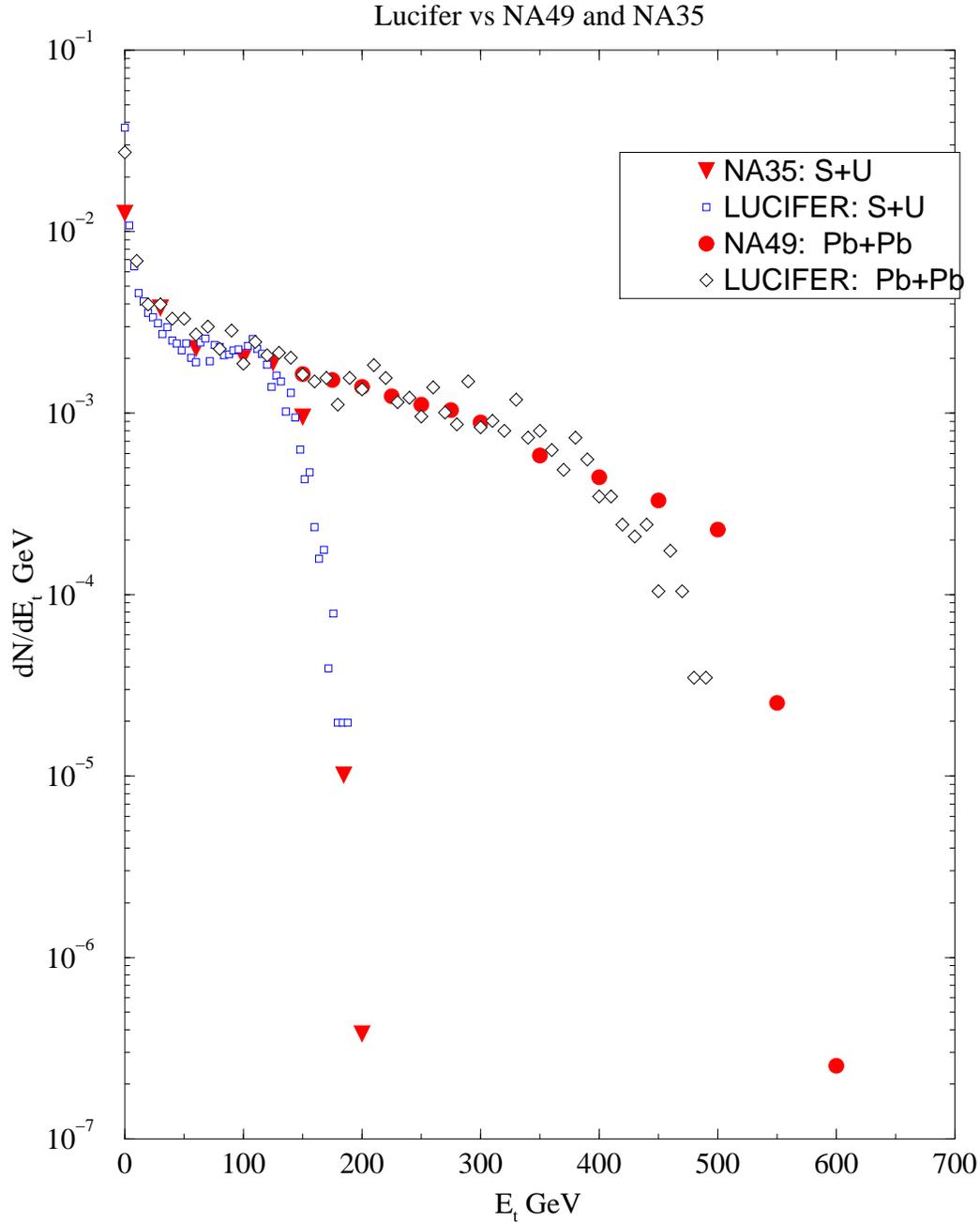}
\hfil}}
\vspace*{10mm}
\caption[]{Transverse energy distributions fro LUCIFER compared to experiment
(NA49) for all charges of hadrons. The purely neutral energy inferred from
this figure should give an upper limit for that seen in the more peripheral
cut used (Figure 1 in the present note.) for NA50.}
\label{fig:two}
\end{figure}
\clearpage 

\begin{figure}
\vbox{\hbox to\hsize{\hfil
\epsfxsize=6.4truein\epsffile[24 85 577 736]{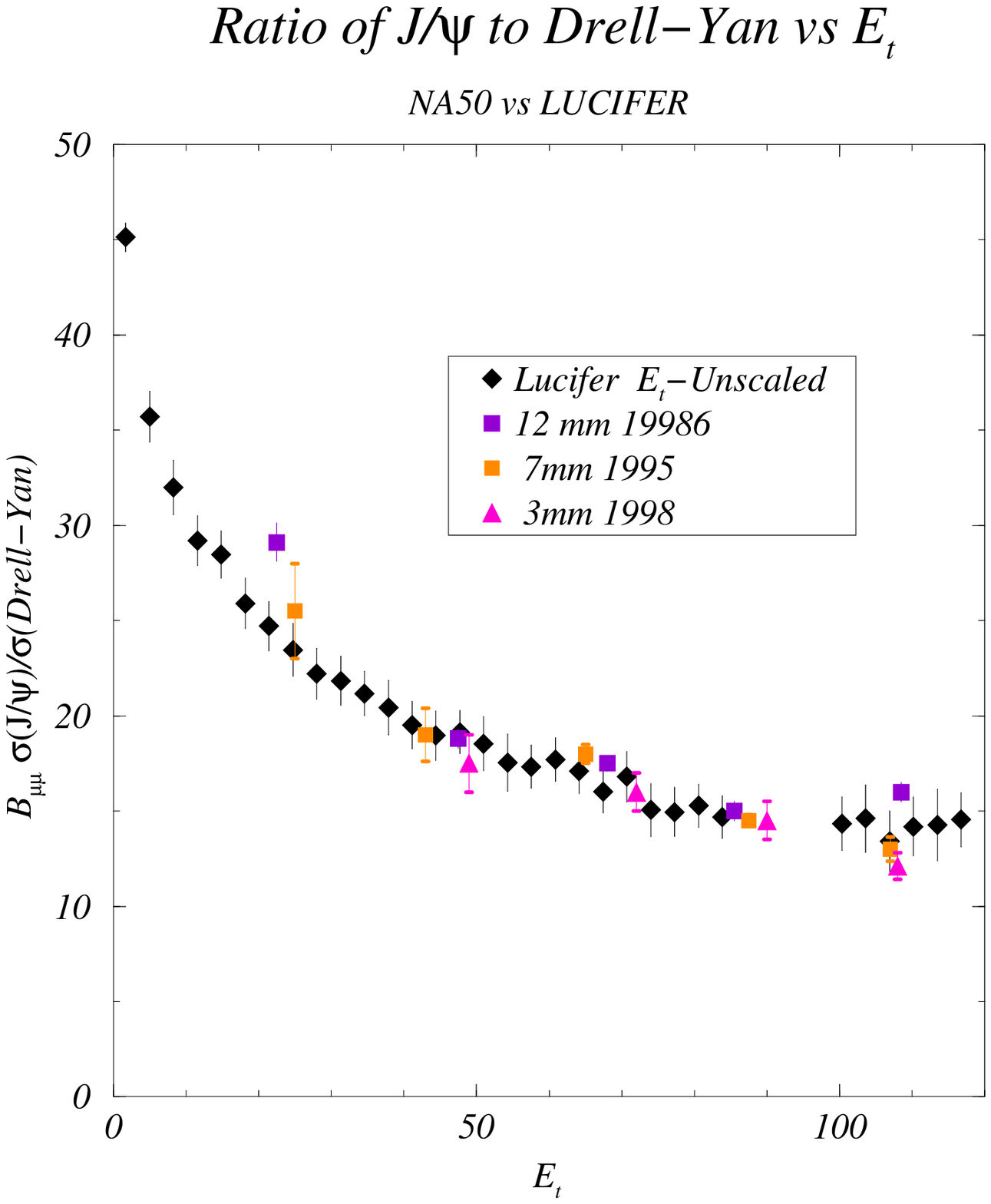}
\hfil}}
\caption[]{Unscaled transverse energy ($E_t$) plot of LUCIFER J/$\psi$ to
Drell-Yan ratio vs the NA50 results presented at QM'99 (Kluberg Figure 57).
Clearly with the two $E_t$ scales more or less in agreement the earlier
description of the J/$\psi$ suppression by LUCIFER \cite{Kahana99,Kahana98}
is reproduced. A slightly revised value (by $\sim 5$\%) for the Drell-Yan
denominator is used here, consistent with uncertainties in the
elementary cross-sections.}
\label{fig:three}
\end{figure}

\clearpage

\end{document}